\documentclass[doublecol]{epl2} 
\usepackage{amsfonts}
\usepackage{amsmath}
\usepackage{amssymb}
\usepackage{graphicx}
\title{Finite element method for obtaining the regularized photon Green function in lossy material}

\author{Meng Tian\inst{1} \and Yong-Gang Huang\inst{1}\thanks{E-mail: \email{huang122012@163.com}} \and Sha-Sha Wen\inst{1}\and Hong Yang\inst{1}\and Xiao-Yun Wang\inst{1} \and Jin-Zhang Peng\inst{1}\and He-Ping Zhao\inst{1} }
\shortauthor{F. Author \etal}

\institute{
  \inst{1} College of Physical Science and Mechanical Engineering, Jishou University,Jishou 416000, China
}

\abstract{
Photon Green function (GF) is the vital and most decisive factor in the field of quantum light-matter interaction. It is divergent with two equal space
arguments in arbitrary-shaped lossy structure and should be regularized. We introduce a finite element method for calculating the regularized GF. It is expressed by the averaged radiation electric field over the
finite-size of the photon emitter. For emitter located in homogeneous lossy material,
excellent agreement with the analytical results is found for both real cavity
model and virtual cavity model. For emitter located in a metal nano-sphere, the
regularized scattered GF, which is the difference between the
regularized GF and the analytical regularized one in homogeneous
space, agrees well with the analytical scattered GF.
}

\begin{document}

\maketitle

\section{Introduction}
According to quantum electrodynamics, the modifications of the spontaneous
emission rate and the energy level shift of a quantum emitter in arbitrary
structure can be expressed in terms of the classical GF \cite{novotny2006principles,agarwal1974quantum,PhysRevA.68.043816,PhysRevA.70.053823,Buhmann2012Dispersion,Buhmann2012Dispersion2,Fuchs2018Purcell,Chen2015Radiation,Hughes:18,Lu2018Influence,PhysRevLett.118.237401,Zhang2017Sub,Chen2016On}. However,
both the real part and imaginary part of the GF with two equal space
arguments in lossy structure are divergent \cite{Jackson1998Classical}, which lead to unphysical divergent
spontaneous emission rate and energy level shift \cite{Van2012Finite}. Real cavity model or virtual
cavity model is usually adopted ( see Ref. \cite{PhysRevA.60.4094} and references therein ), where the emitter is assumed to be in a small lossless cavity. Thus,
the local field seen by the emitter is different from the macroscopic field.
For the real cavity model, the lossless small cavity introduces a dielectric
mismatch, which leads to additional scattering. In this case, the scattered GF
can be used to express the spontaneous emission rate and the energy level
shift for a point emitter. But for larger photon emitters such as quantum dots
and macromolecules, regularized GF, which is the averaged GF over the photon
emitter, may be more proper \cite{wulixuebao,Yaghjian1980Electric,Martin1998Electromagnetic,PhysRevE.70.036606,PhysRevA.60.1590}. For the virtual cavity model, it is assumed that the field
outside the fictitious cavity is the same as there is no cavity. The local
field is the sum of the average macroscopic field and the internal field,
where the internal field is the difference between the actual contribution and
the averaged one of the molecules in the virtual cavity. For the
Clausius-Mosotti type, the actual contribution is thought to be zero and the
local field is expressed by the average macroscopic field \cite{Jackson1998Classical,PhysRevA.60.4094}. So, singular behavior remains and regularization is also required \cite{PhysRevA.60.4094}.

According to the field theory \cite{novotny2006principles}, GF is expressed by the electric field of a
radiating electric point dipole and can be obtained by a number of ways ( for example,
see a recent review \cite{Gallinet2015Numerical} ). For
arbitrary nano-structure, finite difference time domain (FDTD) method and
finite element method (FEM) are two popular numerical methods. It has been
shown that FDTD method can be used to calculate the regularized GF \cite{Van2012Finite}. The size of the mesh grid should be smaller than
that of the regularization volume, which is nearly the size of the photon
emitter and is usually sub-nanometer. According to FDTD method, the smaller the mesh
grid size is, the longer the computation time is. In addition, staircase
effect can not be avoided. But for the FEM, flexible discretization strategy
and the use of basis function enable one to account for extremely complex and sophisticated nano-structure, which are designed to manipulate the light-matter interaction\cite{Li2015Optics,Burresi2015Complex}. Recently, FEM is used to calculate the scattered
GF \cite{Zhao:18} and the regularized GF \cite{wulixuebao} for emitter located around a plasmonic nano-structure. However, its performance for emitter located in lossy material is not clear.

In this work, we demonstrate that FEM can be accurately applied to compute the regularized GF. Analytical results for emitter located in homogeneous lossy
material and metal nano-sphere are presented as a reference. Both real cavity
model and virtual cavity model are taken into account. We first describe the
model and present the numerical method. Next, we make a comparison between
the results by FEM and the analytical ones.

\section{Model and Method}

As a demonstration, the model and parameters are the same as those in Ref. \cite{Van2012Finite}. The lossy
material is chosen to be silver ( Ag ) with a permittivity given by the Drude model
$\varepsilon_{2}=\varepsilon_{\infty}-\omega_{p}^{2}/(\omega^{2}-i\gamma
\omega)$, where $\varepsilon_{\infty}=6$, $\omega_{p}=7.89eV$ and
$\gamma=0.051eV$. Figure 1 shows the schematics of the geometries. For emitter located in the homogeneous lossy material, virtual cavity model and real cavity model are shown in Fig. 1(a) and 1(b), respectively. For nonhomogeneous nano-structures, we choose a nano-sphere with radius $r_{a}$ as shown in Fig. 1(c) and 1(d) for virtual and real cavity models, respectively. The regularization volume is chosen to be a small sphere around the emitter with
radius $a$ and permittivity $\varepsilon_{1}$. In these cases, analytical
regularized GF can be obtained. For simplicity, we set $\varepsilon
_{1}=\varepsilon_{2}$ and $\varepsilon_{1}=1$ for the virtual cavity model and the real cavity model respectively, although $\varepsilon_{1}$ should be the permittivity of the emitter. It should be noted that the regularization
volume can be of any other shape, such as cubic or pyramid. In addition, the
permittivity $\varepsilon_{1}$ can be of any realistic value for the quantum emitter.

\begin{figure}[htbp]
\centering
\includegraphics[width=8cm]{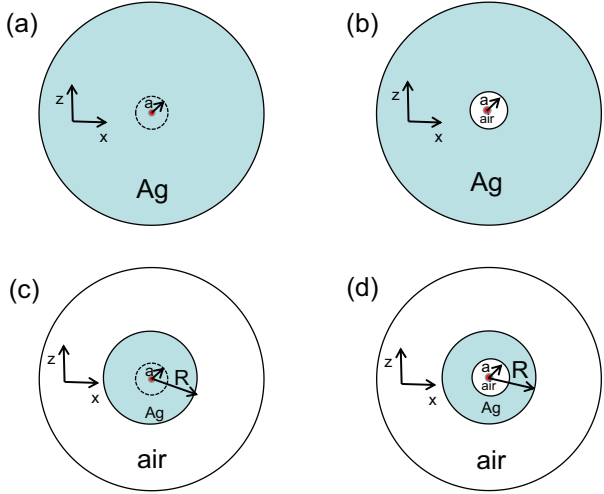}
\caption{Schematics for the regularized GF averaged over a sphere with radius $a$, for (a) homogeneous material and (c) an Ag nano-sphere. The permittivity is assumed to be the same as the lossy material around for both virtual cavity models. (b) and (d) are the real cavity models where the permittivity is set to $\varepsilon_{1}=1$.}
\label{fig1}
\end{figure}

Theoretically, the photon GF satisfies \cite{PhysRevB.85.075303}
\begin{equation*}
\nabla\times\nabla\times\mathbf{G}(r;r_{0};\omega)-\varepsilon(r,\omega
)\frac{\omega^{2}}{c^{2}}\mathbf{G}(r;r_{0};\omega)=\frac{\omega^{2}}{c^{2}%
}\mathbf{I}\delta(r-r_{0}),
\end{equation*}
where $\mathbf{I}$ is the unit dyadic. Its component can be expressed by the
radiation electric field at $r$ of a radiating point dipole $\mathbf{d}$ at
$r_{0}$. Explicitly, it is $E(r)=\mathbf{G}(r;r_{0};\omega)\cdot
\mathbf{d/}\varepsilon_{0}$. However, for $r\rightarrow r_{0}$ in the source
region, the electric field $E(r)$ is divergent and regularization is required.
The regularized electric field can be written as
\begin{equation}
E^{reg}(r_{0})=\frac{1}{\Delta V}\int_{\Delta V}E(r)dr,
\label{regularized field}%
\end{equation}
with $E(r)=\frac{1}{\varepsilon_{0}}[\int_{\Delta V}\mathbf{G}_{s}%
(r;r_{0};\omega)\cdot P(r_{0};\omega)dr+\lim_{V_{\delta}\rightarrow0}%
\int_{\Delta V-V_{\delta}}\mathbf{G}_{0}(r;r_{0};\omega)\cdot P(r_{0}%
;\omega)dr-\frac{L\cdot P(r;\omega)}{\varepsilon_{B}}]$ \cite{Yaghjian1980Electric}. Here, the GF is
decomposed into a homogeneous part $\mathbf{G}_{0}(r;r_{0};\omega)$ and a scattered part $\mathbf{G}_{s}(r;r_{0}%
;\omega)$, $\mathbf{G}%
(r;r_{0};\omega)=\mathbf{G}_{0}(r;r_{0};\omega)+\mathbf{G}_{s}(r;r_{0}%
;\omega)$. A \textquotedblleft principal volume\textquotedblright\ $V_{\delta
}$, which excludes the singularity of the homogeneous GF $\mathbf{G}%
_{0}(r;r_{0};\omega)$, is an infinitesimally small volume around $r_{0}$ and
tends toward 0. Thus, the regularized field in Eq. (\ref{regularized field}) can be written as

\begin{equation}
E^{reg}(r_{0})=\frac{\mathbf{G}^{reg}(r_{0};r_{0};\omega)\cdot\mathbf{d}%
}{\varepsilon_{0}}, \label{reguEToG}%
\end{equation}
where the regularized GF is%
\begin{equation}
\mathbf{G}^{reg}(r_{0};r_{0};\omega)=\mathbf{G}_{s}^{reg}(r_{0};r_{0}%
;\omega)+\frac{M}{\Delta V}-\frac{L}{\Delta V}.\label{regularizedGF}%
\end{equation}
Here the regularized scattered GF is $\mathbf{G}_{s}^{reg}(r_{0};r_{0}%
;\omega)=\int_{\Delta V}\mathbf{G}_{s}(r;r_{0};\omega)dr/\Delta V$. For cubic
or sphere principal volume, the source term is $L=I/3\varepsilon_{b}$. The
self-term $M$ can be numerically calculated for cubic volume \cite{PhysRevE.70.036606}. But for sphere principal volume of radius $a$, it can be
analytically obtained and reads $M=2((1-ika)e^{ika}-1)I/3\varepsilon_{b}$,
where $k=\sqrt{\varepsilon_{b}}\omega/c$ and $\varepsilon_{b}$ is the relative
permittivity in the principal volume $V_{\delta}$.

Numerically, the regularized GF $\mathbf{G}^{reg}(r_{0};r_{0};\omega)$ can be
obtained from Eq. (\ref{reguEToG}), where the regularized electric field
$E^{reg}(r_{0})$ is calculated by first evaluating the radiation field $E(r)$
of a point dipole located at $r_{0}$ and then averaging it over the
regularization volume $\Delta V$ (see Eq. (\ref{regularized field})). Thus,
any modeling approach that can calculate the radiation field $E(r)$ of a point
dipole can be applied. FEM is one of the most proper methods for this problem,
especially for emitter located in nano-structure with arbitrary shape. The wave
equation is transformed into its variational form, where point dipole can be
treated properly \cite{PhysRevB.81.125431,Gallinet2015Numerical}. By the full-wave three-dimensional simulations based on FEM (COMSOL
Multiphysics), the radiating electric field $E(r)$ of a point dipole has been
calculated \cite{Zhao:18,wulixuebao,OptExpress2127371}. Although
the field at the source location depends on the mesh grid, we have shown that
both the scattered field and the regularized one are independent and agree
well with the analytic for emitter located in nonlossy
material \cite{Zhao:18,wulixuebao}. In this work, we develop this method for emitter located in lossy material.

For arbitrary nano-structure, the procedure presented in
Ref. \cite{Zhao:18,wulixuebao} can be used. But for the models considered in this work, an axial symmetry can be used to reduce the computational cost with a so-called 2.5D
implementation if we are interested in the radial component of the regularized
GF $\mathbf{G}_{nn}^{reg}(r_{0};r_{0};\omega)=n\cdot\mathbf{G}^{reg}%
(r_{0};r_{0};\omega)\cdot n$ \cite{Ciraci:13}. In the
following, let us begin with a summary of the main aspects of the modeling technique.

The wave equation with a harmonic source term (time dependence in $e^{-i\omega
t}$ ) $J(r,\omega)\ $reads%

\begin{equation}
\lbrack\nabla\times\frac{1}{\mu_{r}}\nabla\times-k_{0}^{2}\varepsilon
(r)]E(r,\omega)-i\omega\mu_{0}J(r,\omega)=0. \label{waveEq}%
\end{equation}

In the 2.5D formulation, the fields are decomposed in terms of the azimuthal
mode number $m$. For example, the electric field is written:
\begin{equation*}
E(\rho,\phi,z)=\sum_{m}E^{(m)}(\rho,z)e^{im\phi}.
\end{equation*}

If the system is rotationally symmetric ( $\varepsilon(r)$ and $\mu_{r}$ are
independent of $\phi$ ), each cylindrical harmonic propagates independently.
Thus, for each mode number $m$, the wave equation in Eq. (\ref{waveEq}) reads%

\begin{equation}
\lbrack\nabla^{(m)}\times\frac{1}{\mu_{r}}\nabla^{(m)}\times-k_{0}%
^{2}\varepsilon(\rho,z)]E^{(m)}(\rho,z)-i\omega\mu_{0}J^{(m)}(\rho,z)=0,
\label{25DwaveEq}%
\end{equation}
where the operator $\nabla^{(m)}\times$ is obtained from the curl operator by
substituting derivatives with respect to $\phi$ with a factor $im$ in
cylindrical coordinates. By multiplying the above equation with the test
function $F^{(m)}(\rho,z)$ and integrating by parts, we get the weak form for
the electric field
\begin{equation}
\begin{aligned}
2\pi\int_{S}[(\frac{1}{\mu_{r}}\nabla^{(m)}\times E^{(m)}(\rho,z))\cdot
(\nabla^{(m)}\times F^{(m)}(\rho,z))- \\
(k_{0}^{2}\varepsilon(\rho,z)+i\omega
\mu_{0}J^{(m)}(\rho,z))\cdot F^{(m)}(\rho,z)]\rho d\rho
dz=0.\label{25Dweakform}
\end{aligned}
\end{equation}
For our models, point dipole is located on the z-axis with $p=de_{z}$. The
current density is $J=-i\omega de_{z}\delta(z-z0)$ which only contains the
$m=0$ term in the cylindrical harmonics decomposition. Note that this equation
is solved on the two-dimensional cross-section of the simulation domain.

In this work, we numerically solve Eq. (\ref{25Dweakform}) by using the RF module of COMSOL
Multiphysics. 2D axisymmetric is employed. Perfectly matched layers (PML) are
used to simulate the unbounded or infinite domains. The size of the simulation
domain is of the order of $2\lambda^{2}$. A nonuniform mesh is employed. Since
the fields change dramatic around the source and volumn average is requried
(Eq. (\ref{regularized field})), a maximum mesh size of $0.1nm$ is used around
the source.  We have checked that these parameters provide accurate numerical convergence.

To conclude this section, we first solve Eq. (\ref{25Dweakform}) for point
dipole source to obtain the radiating electric field $E(r)$. Then,
$E^{reg}(r_{0})$ can be evaluated from Eq. (\ref{regularized field}). By Eq.
(\ref{reguEToG}), the component of the regularized GF can be obtained easily.
For arbitrary nano-structure, the regularized scattered GF $\mathbf{G}_{s}%
^{reg}(r_{0};r_{0};\omega)$ can be obtained by subtracting the analytical regularized
homogeneous GF $\mathbf{G}_{0}^{reg}(r_{0};r_{0};\omega)=\frac{M}{\Delta V}-\frac{L}{\Delta V}$ ( the last two terms on the right hand of Eq.
(\ref{regularizedGF}) ) from the regularized GF $\mathbf{G}^{reg}(r_{0}%
;r_{0};\omega)$.

\section{Numerical results}

We first investigate the homogeneous case (models shown in Fig. 1(a) and Fig.
1(b)), where analytical regularized GF can be found. For the virtual cavity
model (Fig. 1(a)), the analytical regularized GF is $\mathbf{G}_{0}^{reg}(r_{0};r_{0};\omega)=\frac{M}{\Delta V}-\frac{L}{\Delta V}$ which can be obtained from Eq.
(\ref{regularizedGF}) with $\mathbf{G}_{s}^{reg}(r_{0};r_{0};\omega)=0$. But
for real cavity model ( Fig. 1(b) ), the regularized scattered GF $\mathbf{G}%
_{s}^{reg}(r_{0};r_{0};\omega)$ is replaced by the scattered GF $\mathbf{G}%
_{s}(r_{0};r_{0};\omega)$ with the assumption that $\mathbf{G}_{s}%
(r;r_{0};\omega)\approx\mathbf{G}_{s}(r_{0};r_{0};\omega)$ for $r$ around
$r_{0}$. In addition, $\mathbf{G}_{s}(r_{0};r_{0};\omega)$ can be rigorously obtained from
the Mie theory \cite{Li1994Electromagnetic,PhysRevA.63.053811,PhysRevB.85.075303}.

\begin{figure}[htbp]
\centering
\includegraphics[width=4cm]{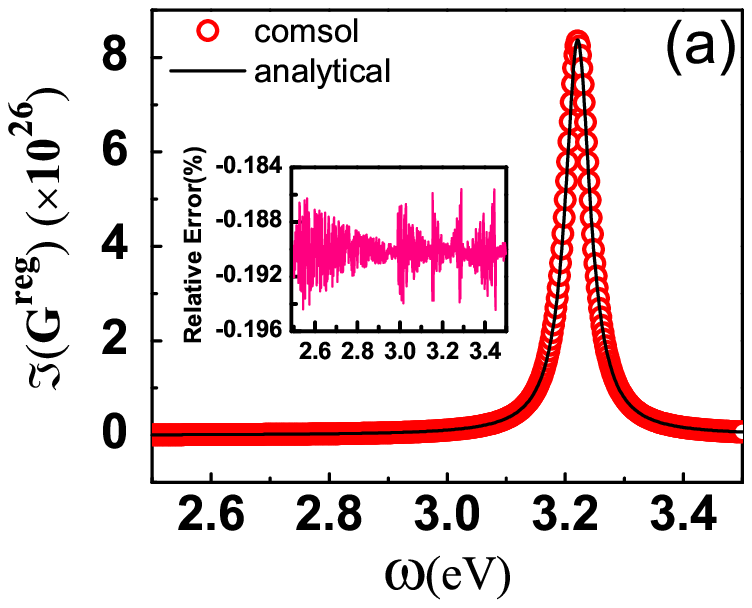}
\includegraphics[width=4cm]{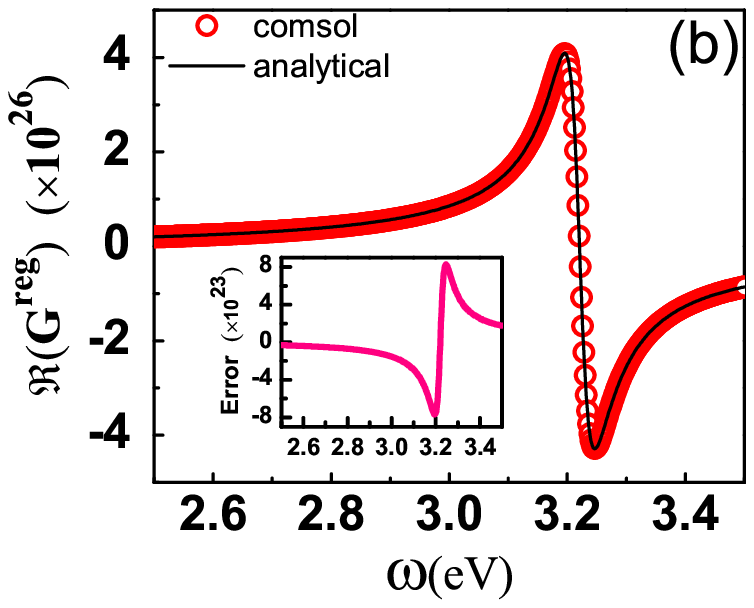}
\includegraphics[width=4cm]{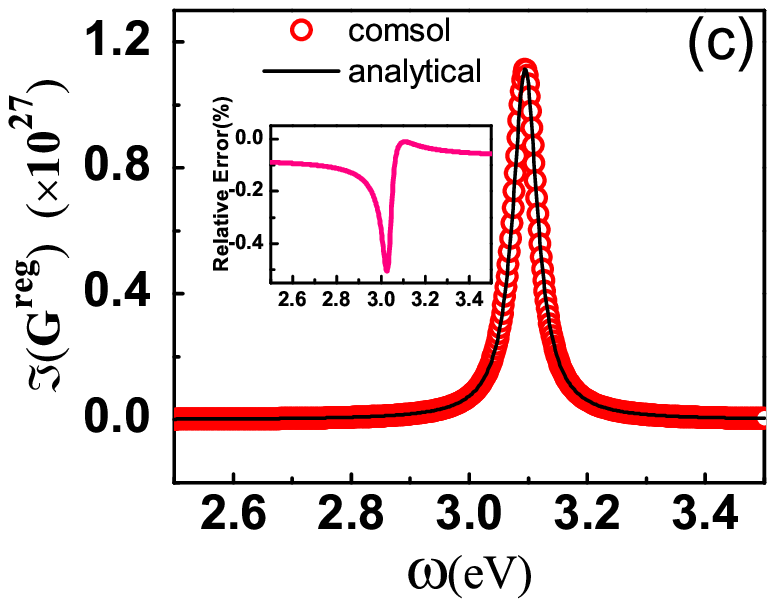}
\includegraphics[width=4cm]{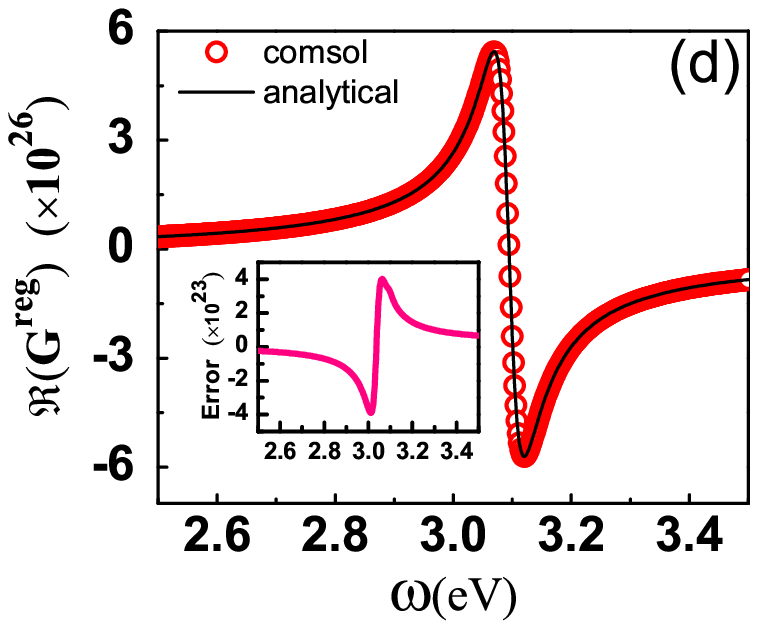}
\caption{Regularized GF for homogeneous lossy material with $a=1nm$. (a) and (b) are the imaginary part and the real part for the virtual cavity model. (c) and (d) correspond to the real cavity model. Red circles are results from the FEM and black lines are for the analytic. Their differences are shown in the insets. The relative errors are defined by the ratio of the differences to the analytical ones.}
\label{fig2}
\end{figure}
Results based on FEM and analytic for the regularized GF are demonstrated in Fig. 2. Here, the
radius for the regularized sphere is $a=1nm$. We find that numerical
results agree well with the analytical solutions, both for the virtual cavity
model (imaginary part shown in Fig. 2(a) and the real part in Fig. 2(b)) and for the real cavity model (Fig. 2(c) and Fig. 2(d) for imaginary part and real part respectively). The relative errors for the imaginary part (see the insets in Fig. 2(a) and Fig. 2(c)) are less than $0.5 \%$ over a wide frequency range. In Fig. 2(a), the large peak at $3.22098eV$ corresponds to the real part of the root for $\varepsilon=0$. This is slightly different from $3.23eV$ obtained by FDTD in Ref. \cite{Van2012Finite}. But for Fig. 2(c), the large peak changes to $3.09459eV$, which corresponds to the real part of the zeros for the denominator $C_N^{22}$ with $n=1$ ( see Eq. ( 26b )in Ref. \cite{Li1994Electromagnetic}). For the real part, errors defined as the difference between the
numerical results and the analytical ones are two orders lower than their values (see the insets in Fig. 2(b) and Fig. 2(d)). These results demonstrate that FEM is accurate and can be applied to investigate the renormalized GF in homogeneous lossy material for both virtual and real cavity models.

\begin{figure}[htbp]
\centering
\includegraphics[width=4cm]{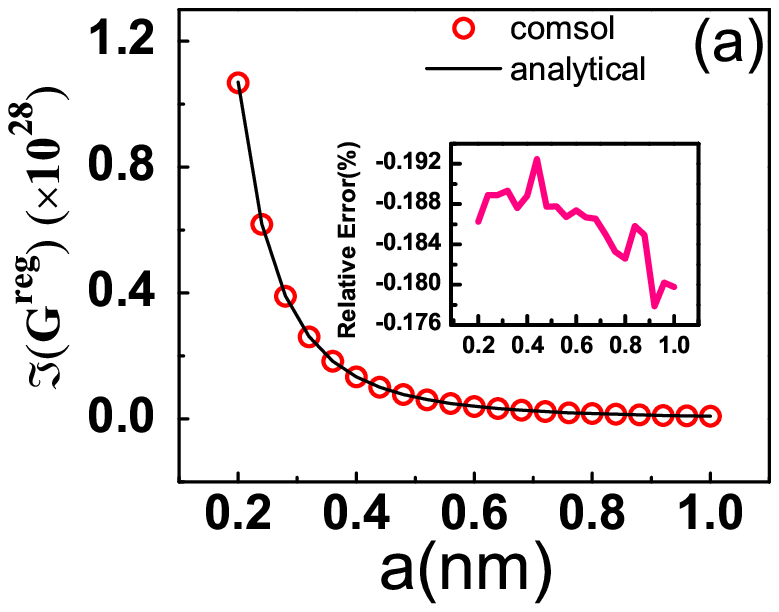}
\includegraphics[width=4cm]{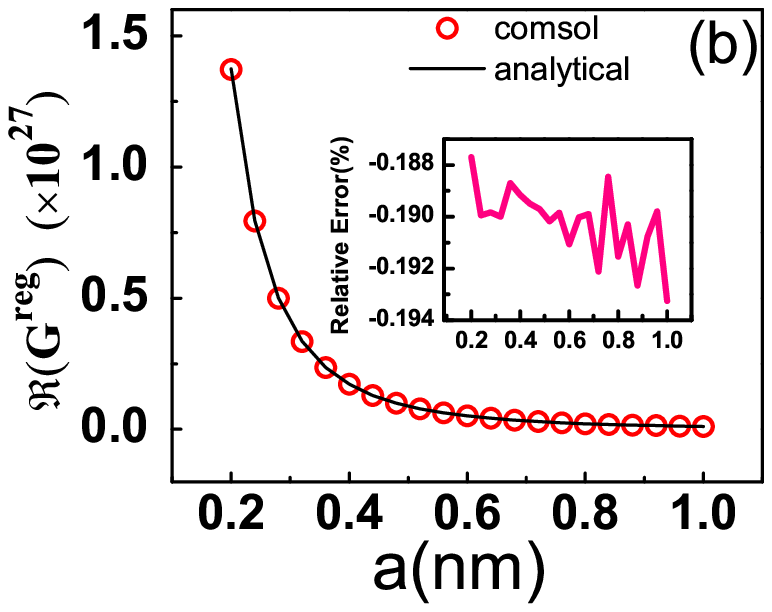}
\includegraphics[width=4cm]{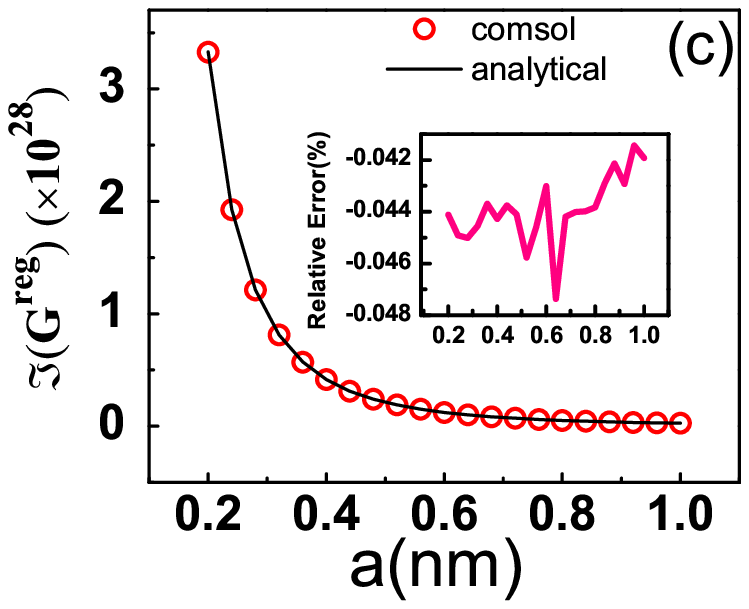}
\includegraphics[width=4cm]{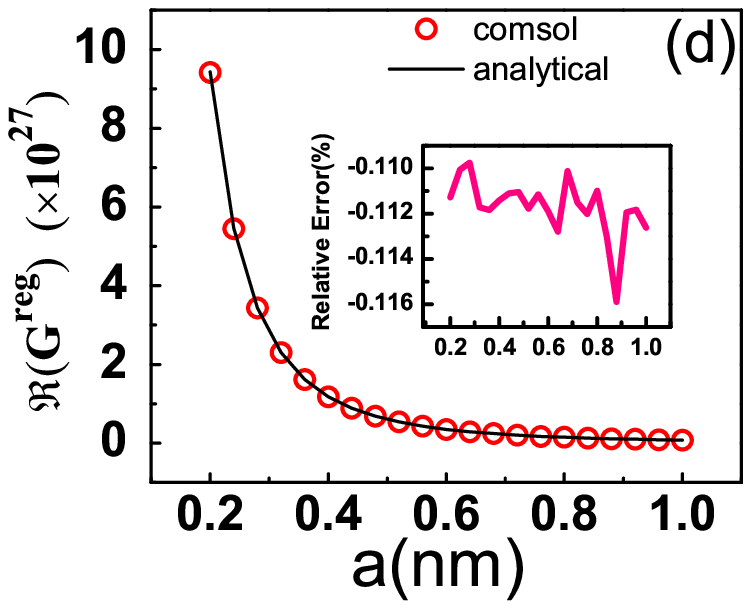}
\caption{Regularized GF as a function of the radius for the renormalization sphere $a$. (a) and (b) are the imaginary part and the real part for the virtual cavity model respectively. (c) and (d) correspond to the real cavity model. Red circles are results from the FEM and black lines are for the analytic with the insets for their relative differences. }
\label{fig3}
\end{figure}

The above results are for $a=1nm$. In Fig. 3, we investigate the effect of
the size for the emitter on the regularized GF. Here, the frequency is set to $\omega=3 eV$ for simplicity. Figure 3(a) and 3(b) are results for the virtual cavity model and Fig. 3(c) and 3(d) are for real cavity model. We observe that numerical results also agree well with the analytic ( to within less than $0.2\%$ ). In addition, all the results decrease as a function proportional to the radius of the small sphere $a^{-3}$. For the virtual cavity model, this can be clearly seen from Eq. (\ref{regularizedGF}) where the first term is zero and the self-term $M$ is about five orders lower than the source term $L$ for the radius considered. For the real cavity model, this is consistent with the result in Ref. \cite{PhysRevA.60.4094}( for example see Eq. (52) therein for extremely small $a$) , which is attributed to the character of the dipole-dipole interaction between the atom and the medium.
\begin{figure}[htbp]
\centering
\includegraphics[width=4cm]{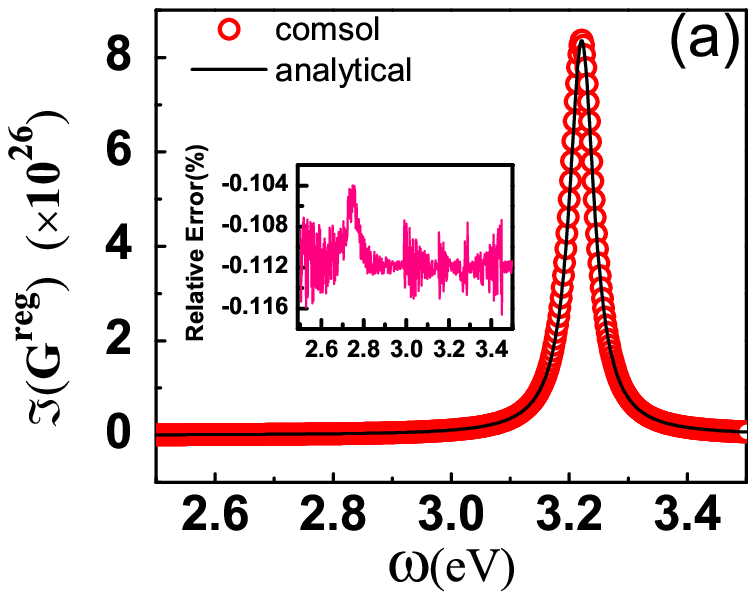}
\includegraphics[width=4cm]{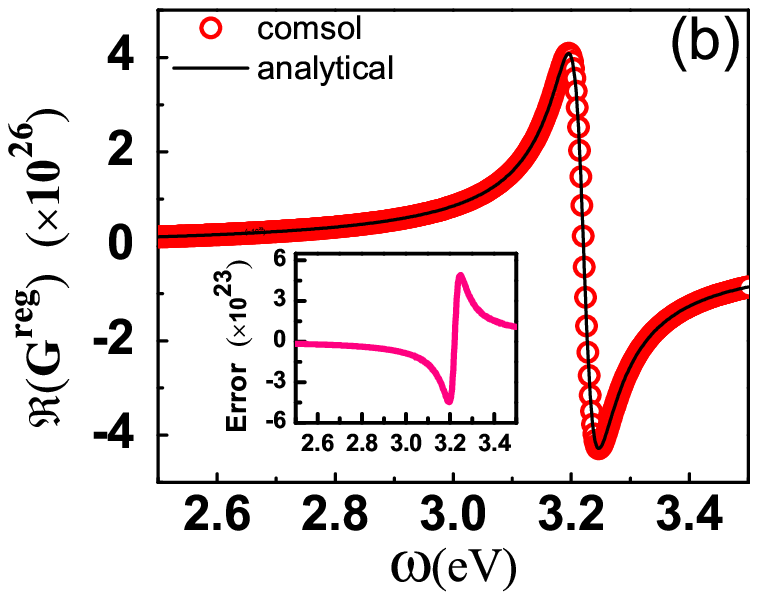}
\includegraphics[width=4cm]{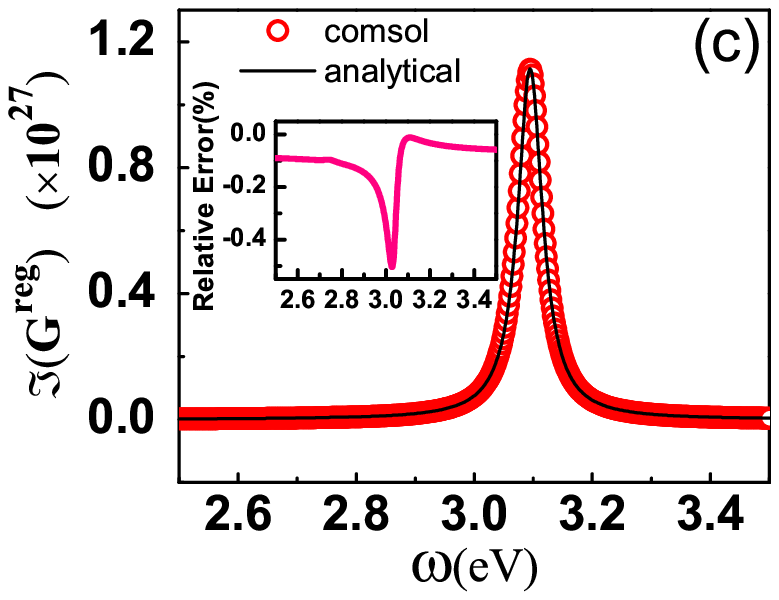}
\includegraphics[width=4cm]{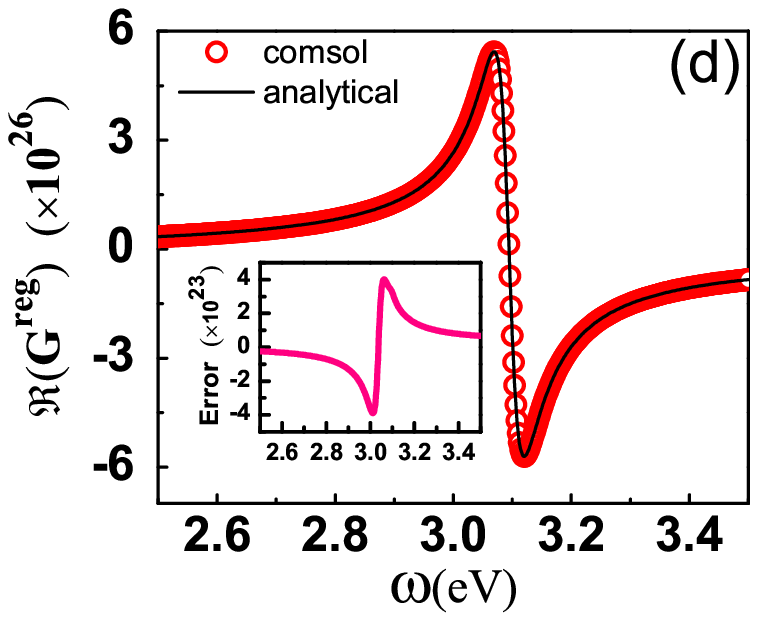}
\caption{Regularized GF for emitter located at the center of a silver nanosphere with $a=1nm$. (a) and (b) are the imaginary part and the real part for the virtual cavity model. (c) and (d) correspond to the real cavity model. Red circles are results from the FEM and black lines are for the analytic. Their differences are shown in the insets. The relative errors are defined by the ratio of their differences to the analytical ones.}
\label{fig4}
\end{figure}
Then, we turn to the case for emitter located in a silver nanosphere (shown in Fig. 1(c) for virtual cavity model and Fig. 1(d) for real cavity model). For both the virtual cavity model and the real cavity model, the analytical results are obtained from Eq. (\ref{regularizedGF}) with the assumption that the scattered GF $\mathbf{G}_{s}(r;r_{0};\omega)$ varies slowly with the space argument $r$. Thus, the regularized scattered GF $\mathbf{G}_{s}^{reg}(r_{0};r_{0};\omega)$ is replaced by the scattered GF $\mathbf{G}_{s}(r_{0};r_{0};\omega)$, which can be calculated by the technique in Ref. \cite{Li1994Electromagnetic}. From Fig. 4, we find that numerical results also agree well with the analytical ones and their differences shown in the insets are also very small.
\begin{figure}[hbtp]
\centering
\includegraphics[width=4cm]{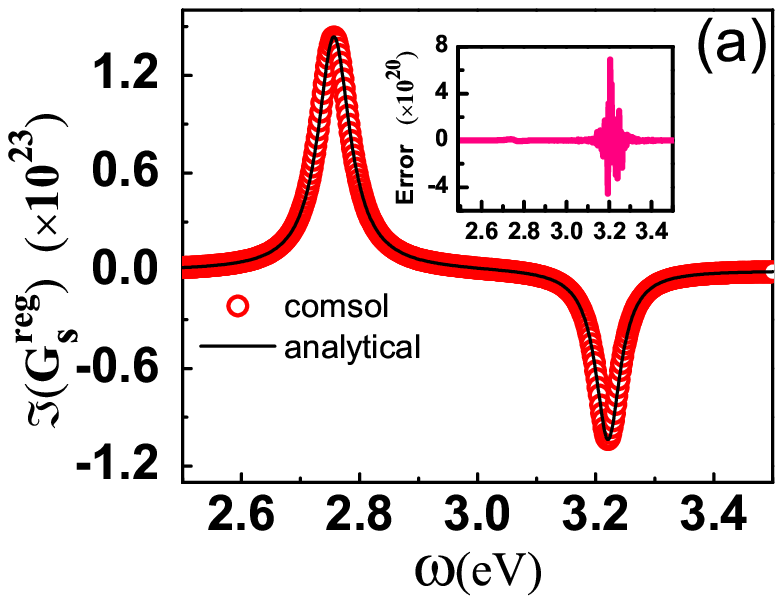}
\includegraphics[width=4cm]{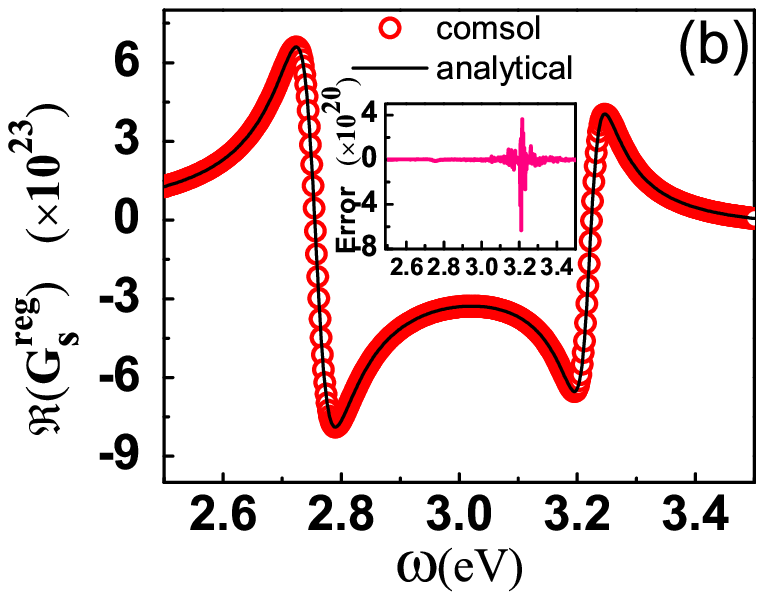}
\includegraphics[width=4cm]{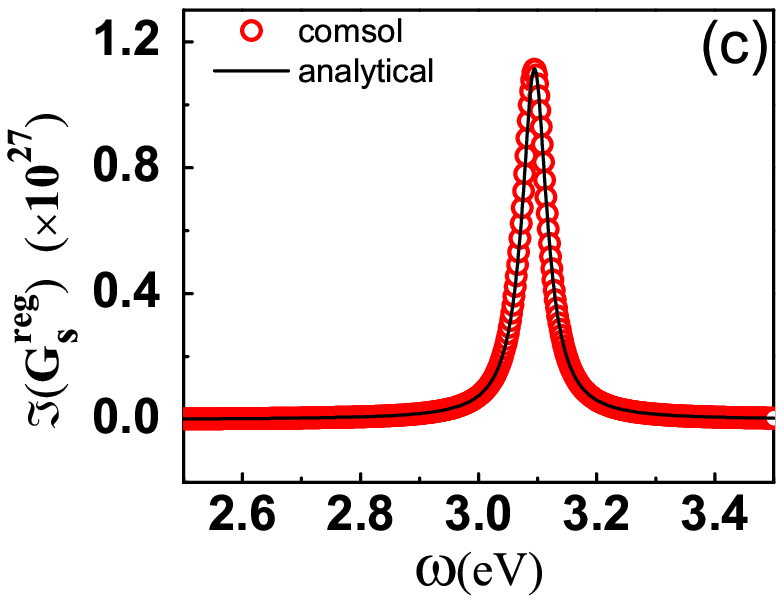}
\includegraphics[width=4cm]{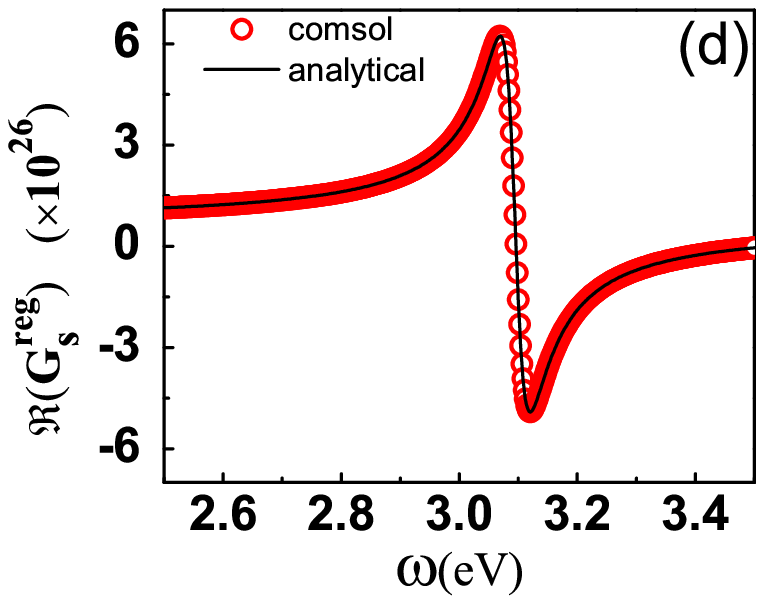}
\includegraphics[width=4cm]{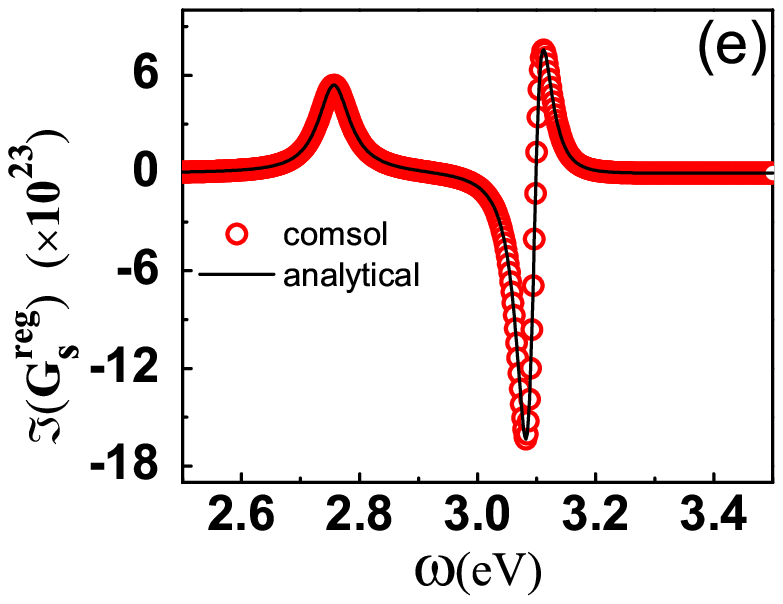}
\includegraphics[width=4cm]{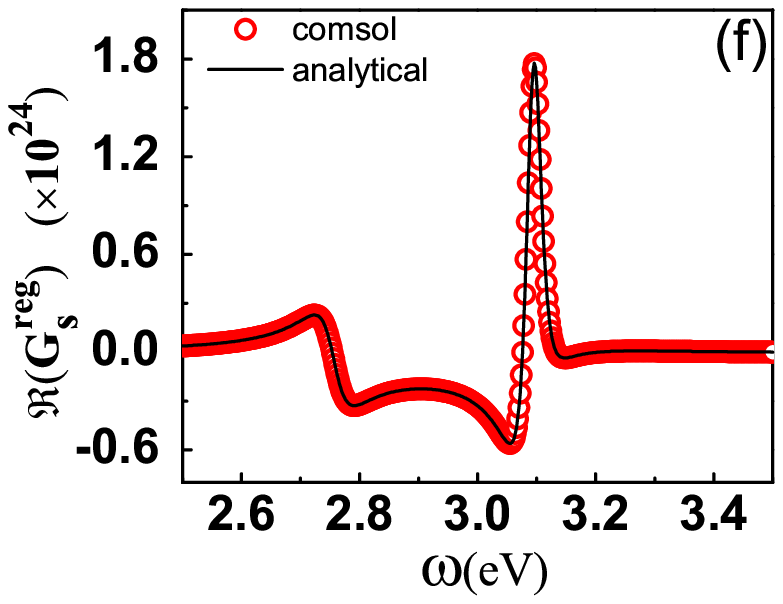}
\caption{Regularized scattered GF for emitter located at the center of a silver nanosphere with $a=1nm$. (a) and (b) are the imaginary part and the real part for the virtual cavity model. (c) and (d) correspond to the real cavity model. (e) and (f) represent the differences between the case for nano-sphere and the homogeneous case. Red circles are results from the FEM and black lines are for the analytic. Their differences are shown in the insets in (a) and (b).}
\label{fig5}
\end{figure}

In addition, the regularized GF for emitter at the center of nano-sphere are similar to those for emitter in homogeneous lossy material. This can be seen by comparing the results in Fig. 4 with those in Fig .2. Thus, the scattering part from the surface of the nano-sphere is far weaker than the homogeneous part. In Fig. 5, we show the results for the scattered GF. We find that numerical results also agree well with the analytical ones. Here, the solid line is the analytical scattered GF $\mathbf{G}_{s}(r_{0};r_{0};\omega)$ and the red circle are for $\mathbf{G}_{s}^{reg}(r_{0};r_{0};\omega)=\mathbf{G}^{reg}(r_{0};r_{0};\omega)-\mathbf{G}_{0}^{reg}(r_{0};r_{0};\omega)$ where the regularized GF $\mathbf{G}^{reg}(r_{0};r_{0};\omega)$ is obtained by FEM as those in Fig. 4 and the regularized homogeneous GF is analytically obtained by $\mathbf{G}_{0}^{reg}(r_{0};r_{0};\omega)=\frac{M}{\Delta V}-\frac{L}{\Delta V}$. Compared to the homogeneous results shown in Fig. 2(a) and 2(b), the magnitudes for the regularized scattered GF are about four orders weaker for the virtual cavity model shown in Fig. 5(a) and 5(b). But for the real cavity model, the regularized scattered GF shown in Fig. 5(c) and Fig. 5(d) looks similar to those shown in Fig. 2(c) and 2(d). Their differences are shown in Fig. 5(e) and Fig. 5(f). Its magnitude is also about three orders lower than the total scattered GF. Thus, for both virtual cavity model and real cavity model, the scattering at the surface of the nano-sphere has little effect. In addition, we find that numerical results agree well with the analytical ones. These results clearly demonstrate that FEM can be able to exactly calculate the regularized scattered GF, although it is the small difference between the total GF and the homogeneous GF.

\section{Conclusions}
 In conclusion, we show that FEM is an efficient and accurate tool for calculating the regularized total GF and scattered GF for both the virtual cavity model and real cavity model in lossy material with arbitrary shape. The regularized total GF is expressed by the averaged electric field, which can be exactly obtained by solving the wave equation through FEM. For the regularized scattered GF, it is expressed by the difference between the total regularized GF and the analytical ones in homogeneous cases. We have confirmed that numerical results from our method agree well with the analytic ones when emitter is located in homogeneous lossy material and a metal nano-sphere.
\acknowledgments
This work was financially supported by the National Natural Science Foundation
of China (Grants No.11464014, 11347215, 11564013, 11464013), Hunan Provincial Innovation Foundation For Postgraduate (Grants No.CX2018B706) and Natural Science Foundation of Hunan, China(Grant No. 2016JJ4073).


\end{document}